\documentclass[conference]{IEEEtran}
\usepackage{cite}
\usepackage{amsmath,amssymb,amsfonts,multirow,array}
\usepackage{algorithmic}
\usepackage{graphicx}
\usepackage{subcaption}
\usepackage{textcomp}
\usepackage{xcolor}
\usepackage[T1]{fontenc}
\def\BibTeX{{\rm B\kern-.05em{\sc i\kern-.025em b}\kern-.08em
    T\kern-.1667em\lower.7ex\hbox{E}\kern-.125emX}}
\begin{document}

\title{Dilated U-net based approach for multichannel speech enhancement from First-Order Ambisonics recordings}

\author{\IEEEauthorblockN{Am\'elie Bosca}
\textit{Orange Labs}\\
Cesson-S\'evign\'e, France \\
amelie.bosca@gmail.com
\and
\IEEEauthorblockN{Alexandre Gu\'erin}
\textit{Orange Labs}\\
Cesson-S\'evign\'e, France \\
alexandre.guerin@orange.com
\and
\IEEEauthorblockN{Laur\'eline Perotin}
\textit{Orange Labs}\\
Cesson-S\'evign\'e, France \\
laureline.perotin@gmail.com
\and
\IEEEauthorblockN{Sr\dj{}an Kiti\'c }
\textit{Orange Labs}\\
Cesson-S\'evign\'e, France \\
srdan.kiticgmail.com
}

\maketitle

\begin{abstract}
We present a CNN architecture for speech enhancement from multichannel first-order Ambisonics mixtures. The data-dependent spatial filters, deduced from a mask-based approach, are used to help an automatic speech recognition engine to face adverse conditions of reverberation and competitive speakers. The mask predictions are provided by a neural network, fed with rough estimations of speech and noise amplitude spectra, under the assumption of known directions of arrival. This study evaluates the replacing of the recurrent LSTM network previously investigated by a convolutive U-net under more stressing conditions with an additional second competitive speaker. We show that, due to more accurate short-term masks prediction, the U-net architecture brings some improvements in terms of word error rate. Moreover, results indicate that the use of dilated convolutive layers is beneficial in difficult situations with two interfering speakers, and/or where the target and interferences are close to each other in terms of the angular distance. Moreover, these results come with a two-fold reduction in the number of parameters.

\end{abstract}


\section{Introduction}
\label{sec:intro}

Speech enhancement is an audio signal processing task which aims to recover a given speech signal from a noisy mixture. This step is very important for applications involving voice commands, especially in far-field conditions where automatic speech recognition (ASR) may suffer from noise and interferences, such as radio, television or other speakers \cite{zhang_deep_2018}.
Enhancement can be achieved by directly predicting the desired signal \cite{stoller_wave-u-net:_2018} or alternatively, by deriving a ratio mask from the magnitude of a time-frequency representation of the mixture \cite{narayanan_ideal_2013}. Recent literature shows that combining spatial information and DNN is highly efficient in enhancing signals corrupted by noise and interference \cite{pertila_distant_2015,heymann_neural_2016,gannot_consolidated_2017,barker_chime_2017}.

Recurrent neural networks (RNNs), in virtue of their memory capacity, have firstly been preferred for time series processing: for waveform generation for instance \cite{tokuday_directly_2015}, or for speech enhancement \cite{huang_deep_2014,weninger_speech_2015}. However, recent works reported that convolutional neural networks (CNNs) were able to perform equally well, if not better, than more complex RNNs, even for audio signals tasks \cite{oord_wavenet:_2016,bai_empirical_2018}. In particular, the U-net architecture, based on an encoding-decoding structure to catch embedded patterns in the input features, has been succesfully applied to audio source separation \cite{stoller_wave-u-net:_2018,jansson_singing_2017}. Very recently, dilated convolution showed interesting results with time series, demonstrating capability to capture long-term information, without increasing network complexity \cite{luo_conv-tasnet:_2019,yazdanbakhsh_multivariate_2019}.

We use spherical microphone arrays, which yield the Ambisonics representation of a sound scene \cite{gerzon_periphony:_1973,pulkki_parametric_2017}. Due to its capacity to efficiently embed 3D-spatial information, this format has been succesfully exploited for source localization \cite{adavanne_sound_2019,perotin_crnn-based_2019} and source separation \cite{perotin_multichannel_2018,epain_independent_2012}. This paper depicts a mask-based Ambisonics speech enhancement system of noisy mixtures based on previous works where masks are estimated via a Long Short-Term Memory (LSTM) recursive network \cite{perotin_multichannel_2018}. The current study investigates the added value of two distinct CNN architectures in place of the LSTM one: a classical U-net and a U-net combined with dilated layers. We generalize their use to the three-speaker scenario.

Section 2 presents the signal model and the Ambisonics speech enhancement system. In Section 3, we detail our U-net architectures. Experiments are presented in Section 4, results in Section 5 and conclusion in Section 6.

\section{Speech enhancement}
 	\subsection{Mixture model}
 	
 	Let us consider a multichannel audio mixture $\textbf{\text{x}}(t,f)$, in the short-time Fourier transform domain, where $t$ and $f$ are the time frame and frequency bin indexes. The mixture $\textbf{\text{x}}(t,f)$ is composed of the target speech signal $\textbf{\text{s}}(t,f)$ to be transcripted by the ASR, and some noise $\textbf{\text{n}}(t,f)$:
 	\begin{equation}
 		\textbf{\text{x}}(t,f) = \textbf{\text{s}}(t,f) + \textbf{\text{n}}(t,f).
 	\end{equation}  
 	 	
The noise component $\textbf{\text{n}}$ gathers some spatially diffuse noise and up to two competitive speakers called interference and named $\textbf{\text{n}}_1$ and $\textbf{\text{n}}_2$. All speakers $s$, $n_1$ and $n_2$ are considered motionless and identified by their spherical coordinates $(\theta,\phi)$, which are supposed to be a priori known.

 	\subsection{First-Order Ambisonics}
	The general Ambisonics format decomposes a 3D sound field on the infinite basis of spherical harmonics functions. A First-Order Ambisonics (FOA) signal is the $1^{st}$ order truncature of such representation and is composed of four channels $(W,X,Y,Z)$: $W$ corresponds to the recording of the sound field by an omnidirectional microphone and $(X,Y,Z)$ are the bidirectionnal captures towards the three cartesian axis. In such formalism, the ideal FOA recording of a plane wave $p(t,f)$ with direction of arrival $(\theta,\phi)$, may be written using the steering vector $\textbf{d}_{\theta,\phi}$:
	\begin{equation}
	\left[
      \begin{array}{c}
        W(t,f)\\
        X(t,f)\\
        Y(t,f)\\
        Z(t,f)\\
     \end{array}
    \right] =
	\textbf{d}_{\theta,\phi} \,\,p(t,f) = 
 \left[
      \begin{array}{c}
        1\\
        \sqrt{3} \,\text{cos}(\theta) \text{cos}(\phi)\\
        \sqrt{3} \,\text{sin}(\theta) \text{cos}(\phi)\\
        \sqrt{3} \,\text{sin}(\phi)\\
     \end{array}
    \right]
    p(t,f)
	\end{equation}

	\subsection{Speech enhancement system}

\begin{figure}[t]
	\hspace{-0.5cm}
	\includegraphics[width=1.05\columnwidth]{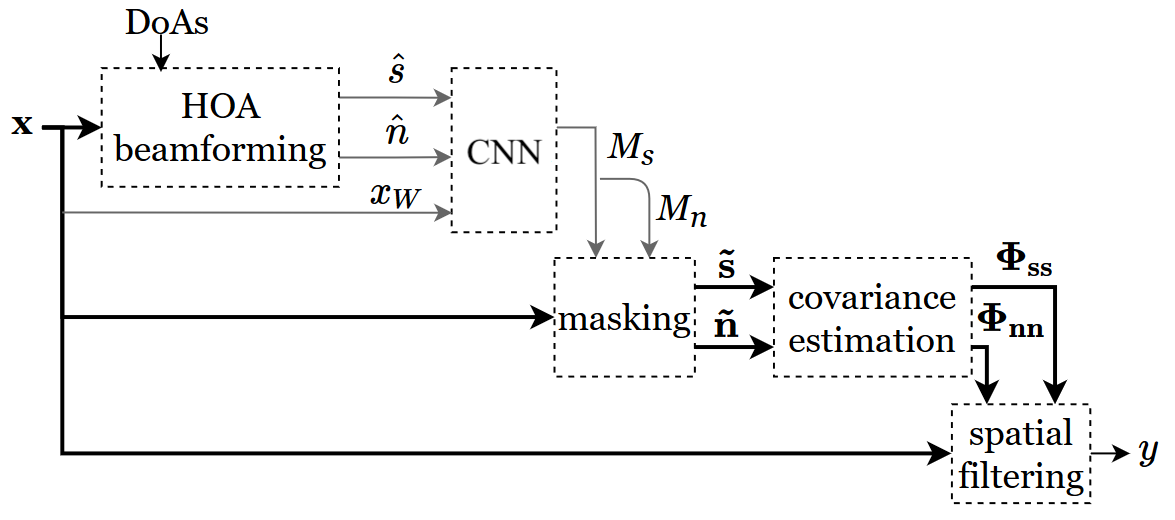}
	\caption{Multichannel speech enhancement system based on masks estimation.}
	\label{fig:structure}
\end{figure}

	The studied enhancement system is based on the same approach described in \cite{perotin_multichannel_2018}, and depicted on Fig. \ref{fig:structure}: first, the neural network provides a real continous mask $M_s(t,f)$ which emphasises the time-frequency points where the target source $s$ is predominant. Then, this mask is used to estimate the covariance matrices of target and noise $\mathbf{\Phi}_{\textbf{ss}}(f)$ and $\mathbf{\Phi}_{\textbf{nn}}(f)$ by temporal integration:	
	\begin{equation}
\left \{
\begin{array}{l}	
\mathbf{\Phi}_{\textbf{ss}}(f) = \frac{1}{T} \sum_{t=0}^{T-1} M_{s}^2 (t,f) \textbf{x} (t,f) \textbf{x}^{H} (t,f) \\
\mathbf{\Phi}_{\textbf{nn}}(f) = \frac{1}{T} \sum_{t=0}^{T-1} (1-M_{s} (t,f))^2 \textbf{x} (t,f) \textbf{x}^{H} (t,f) 

   \end{array}
   \right .
   \label{eq:maskedCovMat}
\end{equation}
where $(.)^H$ is the conjugate transposition and $T$ is the number of frames. These matrices are used to derive a time-independent multichannel Wiener filter (MWF):
\begin{equation}
	\textbf{w}(f) = [\mathbf{\Phi}_{\textbf{ss}}(f) + \mathbf{\Phi}_{\textbf{nn}}(f)]^{-1} \mathbf{\Phi}_{\textbf{ss}}(f) \textbf{u}_1
	\label{eq:MWF}
\end{equation}
	where $\textbf{u}_\mathbf{1}$ is the operator which selects the first column. The estimated target signal $y(t,f)$ is finally deduced by:
	
	\begin{equation}
y(t,f) = \textbf{w}^{\text{H}}(f) \textbf{x}(t,f)
\label{eq:yMWF}
\end{equation}
	
In pratice, this MWF filter is approximated by a rank-1 version using the generalized eigenvalue decomposition (GEVD) and named $\textbf{w}_{\text{GEVD-MWF}}(f)$  \cite{perotin_multichannel_2018}. This filter revealed to be robust to errors in covariance matrix estimation \cite{serizel_low-rank_2014,wang_rank-1_2018}.

	\section{Mask estimation by U-net}
	We proposed a U-net architecture to replace prior work that leveraged an  LSTM-based network to predict ratio masks $M_{s}$ \cite{perotin_multichannel_2018}. As input to our network, we first compute full-band beamformer source estimates based on the ideal FOA plane wave formulation.
	
	\subsection{Which masks $M_{s},M_{n}$ to estimate?}
There exists many different masks emphasizing a specific signal: binary masks, Wiener masks, instantaneous ratio masks. The choice of the masks $M_{s}$ and $M_{n}$ is driven by our use-case: associated with the spatial filter $\textbf{w}_{\text{GEVD-MWF}}(f)$, they shall maximize the ASR performance by  minimizing its WER. 

In our experiments, the instantaneous energy ratio masks $M_{s,n}^{\text{id}}(t,f)$:
\begin{equation}
\left \{
\begin{array}{l} M_{s}^{\text{id}}(t,f) = \frac{\lVert s_{W}(t,f)\rVert ^2}{\lVert s_{W}(t,f)\rVert^2 + \lVert n_{W}(t,f)\rVert^2} \\
M_{n}^{\text{id}}(t,f) = 1-M_{s}^{\text{id}}(t,f)
   \end{array}
\right .
\label{eq:idMask}
\end{equation}
where $\text{.}_{W}$ is the omnidirectional channel, revealed to be good candidates. Indeed, we observed that the enhanced signal $y$ by the combination of eq.(\ref{eq:maskedCovMat},\ref{eq:MWF},\ref{eq:idMask}), produces the same WER as the one enhanced by the oracle multichannel GEVD filter, i.e. using the exact covariance matrices.
	
	\subsection{Input features}
The input features are the same as those of our previous works \cite{perotin_multichannel_2018}, \textit{i.e.} approximations of the magnitude spectra of speech $\hat{s}(t,f) $ and interference $\hat{n}_{1,2}(t,f)$, and the magnitude of the mixture itself $W(t,f)$. These features are ``correlated'' with the quantities necessary to compute the ideal mask $M_s^{id}(f,t) $. The speech and interferences estimations are obtained by full-band ambisonic beamforming, using the a priori known direction of arrival (DOA) of the sources. For instance, the beamformer pointing towards $(\theta_s,\phi_s)$ and canceling $(\theta_{n_{1}},\phi_{n_{1}})$ and $(\theta_{n_{2}},\phi_{n_{2}})$ is given by:
	\begin{equation}
		\textbf{b}_{0}= [\textbf{d}_{\theta_s,\phi_s}~ \textbf{d}_{\theta_{n_{1}},\phi_{n_{1}}}~ \textbf{d}_{\theta_{n_{2}},\phi_{n_{2}}}]^{\dagger}\textbf{u}_\mathbf{1}
		\label{beamformer}
	\end{equation}
	where $(.)^{\dagger}$ is the pseudo-inverse. The estimation of the $i^{\text{th}}$ interference, $i \in [1,2]$ is obtained in the same way by selecting the $(i+1)^{\text{th}}$ column. Beamforming is applied to the mixture using:
		\begin{equation}
		\left\{
		\begin{array}{lcl}
			\hat{s} (t,f) & = & \textbf{b}_{0}^{\text{H}} \textbf{x}(t,f)\\
			\hat{n}_i (t,f) & = & \textbf{b}_{i}^{\text{H}} \textbf{x}(t,f),\;\; \forall i \in [1,2]
		\end{array}
		\right.
		\label{eq:features}
		\end{equation}
	
	In the single interference scenario, inputs of the network are composed of the magnitudes of the spectrograms of the omnidirectional channel of the mixture $|\text{x}_W(t,f)|$, and of the estimations $|\hat{s} (t,f)|$ and $|\hat{n}_1 (t,f)|$ defined by (\ref{eq:features}). The addition of $|\hat{n}_1 (t,f)|$ has shown significant improvements \cite{perotin_multichannel_2018}.

The two interfering speakers case is more complicated. If choosing the interference feature as $|\hat{n}_1 (t,f) + \hat{n}_2 (t,f)|$ seems natural, preliminary results have shown some dramatic drop of performance when using the same network as for the one interfering speaker case. Reason lies in the directivity diagrams of the beamformers which may exhibit some very large sidelobes, especially when sources are close, hence producing amplified output signals. As a result, the standardization applied to the input features is no longer effective. To circumvent this, we chose to apply a  sequence-dependent normalization to each feature before computing its statistics, by dividing each feature $ q=|\hat{s}|,|\hat{n}_1|,|\hat{n}_2|$ in each frequency band by its maximum over the whole input sequence:
\begin{equation}
\tilde{q}(t,f) = \frac{q(t,f)}{\max\limits_{t}q(t,f)}
\end{equation} 
This resulted in improving the overall WER performance by uniformizing the scores for every item of the database, even when beamformers are ill-conditioned.

	\subsection{U-net architecture}
	\label{sec:Unet}
	\begin{figure}

	\includegraphics[scale=0.29]{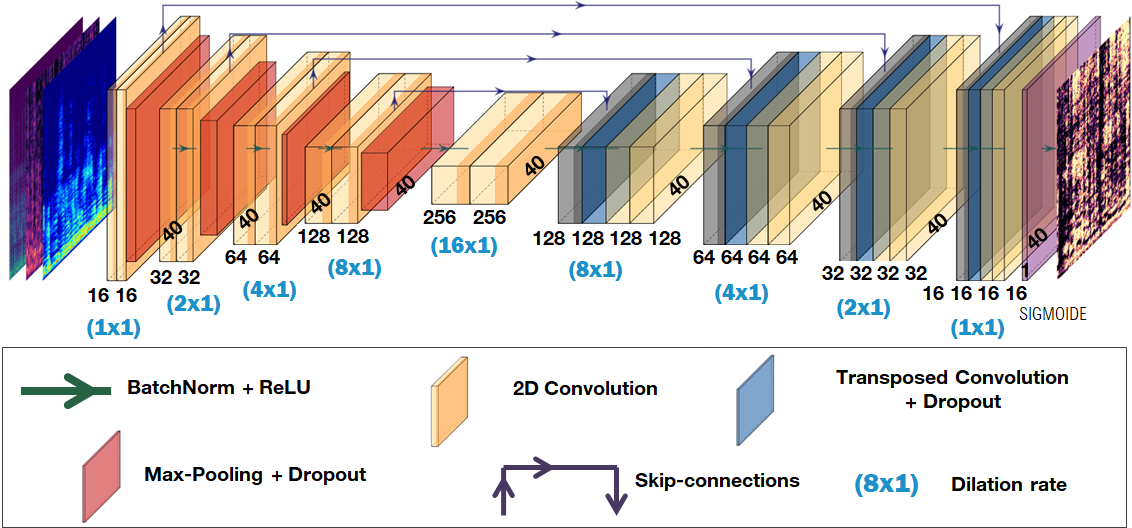}	
	\caption{U-net architecture. Dilation rates for the second version of the network are indicated in blue.}
	\label{Unet}
	\end{figure}	
	
	Our CNN, depicted on Fig. \ref{Unet}, is a U-net composed of nine encoder-decoder blocks. Five encoder blocks provide the compressed representation of the inputs, each block being composed of two 2D convolutions each followed by batch normalization and ReLU activation. After the second convolution, max-pooling is applied in the frequency dimension: this operation may take advantage of speech characteristics (particularly, harmonic structures), to identify the mask patterns highly related to these features. No max-pooling is applied along the time dimension. Indeed, the stationnarity of speech signals, hence of the predicted masks as well, is about 60~ms: with a 30~ms resolution, tests confirmed that applying such max-pooling would result in definitly losing some local patterns (see section \ref{sec:Experiments} for details). The fifth block (the central one) is free from max-pooling. Four decoder blocks re-extend the deep representation: in each block, a transposed convolution layer, performing the deconvolution operation, is concatenated with the output of the encoder block at the same depth using the skip-connections. Two 2D convolutions with batch normalizations and ReLU activations are executed at the end of each decoder block.
	
	To capture local information, the size of the 2D-convolution filters is chosen as (3,3). The number of filters in the first feature map is set to 16 and multiplied by two at each encoding block. Symmetrically, it is divided by two at each decoding block. Max-pooling is chosen of size 2 in the frequency dimension, time dimension remaining unchanged.
	
	The final 2D convolution corresponds to a learned weighted average of the features computed, in order to get the ability of each bin to discriminate the target source between noise and interfering source(s). This layer is followed by a sigmoidal activation to produce the estimated ratio mask.

	\subsection{Dilated layers}
\begin{figure}
	\centering
	\includegraphics[scale=0.15]{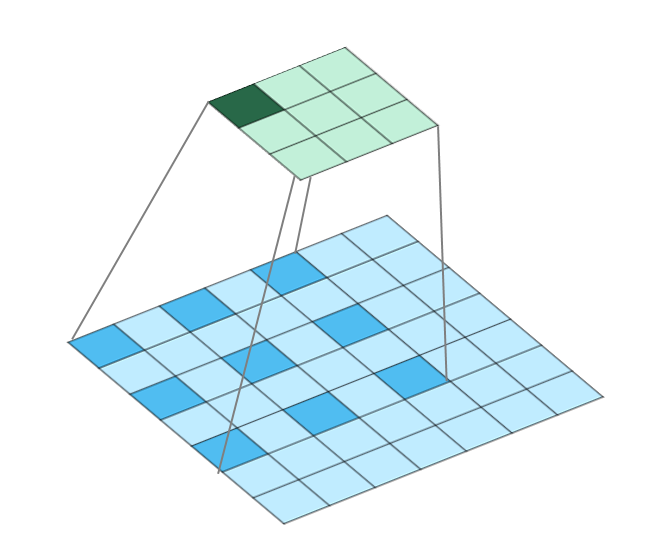}	
	\caption{Dilated Convolution principle with rate 2}
	\label{dilated}
\end{figure}

	In this paper, we also investigate the added value of dilated convolution layers (Fig. \ref{dilated}). Such convolutions are commonly used to increase the receptive field of a neural network, without increasing its complexity: for instance, WaveNet integrates such layers to enlarge the time scope taken into account \cite{oord_wavenet:_2016}. In our use-case, as the target $M^{id}_s(f,t)$ is a short-term mask, increasing the time field does not seem to be of great importance: this has been confirmed by some tests which exhibited some loss of performance when applying dilation in the time dimension. The use of dilated convolutions was rather motivated by ``encouraging'' the network to exploit the harmonic structure of speech signals. Indeed, at a given time, different speech signals have quasi-surely different pitches. Hence, if the frequency resolution is sufficient high, forcing the network to use this spectral structure may reveal beneficial to identify the mask patterns.
	
	In order to compare networks with same complexity, we keep the U-net architecture described in Section \ref{sec:Unet}, and just change one of the two convolution layers of each block to a dilated one. As stated before, dilation is only applied in the frequency dimension. The dilation rate is increased/decreased by a factor of 2 across each consecutive compression/decompression block, as it optimizes the receptive field with respect to the computational efficiency \cite{oord_wavenet:_2016}.

\section{Experiments}
\label{sec:Experiments}

\begin{table*}
	\centering
	\begin{tabular}{|l|c||>{\centering}p{1.7cm}|>{\centering}p{1.7cm}|>{\centering}p{1.7cm}||c|}
		\hline
		\multicolumn{2}{|c||}{} & \multicolumn{3}{c||}{2 speakers: SIR~=~0~dB (SNR=20dB)} & {\centering 3 speakers: SIR~=~6~dB (SNR=20dB)}   \\
		\cline{3-6}
		\multicolumn{2}{|c||}{} & { 25$^\circ$} & {45$^\circ$} & {90$^\circ$} & {[40 ; 50]$^\circ$ }\\
		\hline
		\hline
		\multicolumn{2}{|c||}{Reverberated speech $s_W$} & \multicolumn{4}{c|}{13.8} \\
		\hline
		\multicolumn{2}{|c||}{Mixture $x_W$}  &85.0 &82.4 &78.2 &64.0\\
		\hline
		\multicolumn{2}{|c||}{Ideal mask $M_s^{\text{id}}$}  &20.5 &18.7 &19.3 &19.0\\
		\hline
		\multicolumn{2}{|c||}{Filter from ideal mask} &17.5 &18.2 &15.5 &18.8\\
		\hline
		\multicolumn{2}{|c||}{Beamformer $\hat{s}$} &79.7 &57.9 &25.4 &53.7\\
		\hline
		\hline
		{\multirow{2}{*}{LSTM}} & Mask $M_{s}$  &77.3 &69.5 &59.7 & 65.3\\
		\cline{2-6}
		{} & Filter $\textbf{w}_{\text{GEVD-MWF}}$ &29.2 &23.8 &16.7 & 28.8\\
		\hline
		{\multirow{2}{*}{U-net}} & Mask $M_{s}$ &70.3 & 40.3& 29.2 &58.0\\
		\cline{2-6}
		&Filter $\textbf{w}_{\text{GEVD-MWF}}$ & 23.7&\textbf{18.9} & \textbf{14.9} & 25.7\\
		\hline
		{\multirow{2}{*}{Dilated U-net}}& Mask $M_{s}$ & 63.1& 39.2 &29.7 &60.9 \\
		\cline{2-6}
		& Filter $\textbf{w}_{\text{GEVD-MWF}}$& \textbf{20.8}& \textbf{19.1} &\textbf{15.1} &\textbf{22.5}\\
		\hline
	\end{tabular}
	\caption{WERs (\%) computed on reference signals (top), at the output of each network (`mask') and after the MWF-GEVD (`filter'). The best enhancement systems are shown in bold. \label{tab:WER}}\medbreak
\end{table*}

	In our experiments, speech and noises come from the same datasets as the ones described in \cite{perotin_multichannel_2018} (speech from BREF database \cite{lamel_bref_1991}, noise from \textit{http://freesound.org}), but we used simulated Spatial Room Impulse Responses (SRIRs) instead of real recorded SRIRs for the training dabatase. The main advantage of synthesized SRIRs lies in the huge acoustic diversity of room configurations and dimensions. While such signals are generated from a simplified model of acoustics propagation, recent works we conducted on source localization reveal that neural networks optimized with synthetic SRIRs generalize well to real conditions \cite{perotin_crnn-based_2019}. The SRIRs database for training  comprises rooms of various dimensions, with $RT_{60}$ between 200 and 800~ms. The minimum angle between two sources is set at 25$^\circ$. In the single interference scenario, the signal-to-interference ratio (SIR) is set at 0~dB. With two interferent speakers, the SIR, defined as the ratio of the target to each interferent speaker, is set to 6~dB, to keep the target source predominant in the mixture: simulations showed no advantage in training the network with more adverse conditions. Some diffuse noise - generated in the same way as \cite{perotin_multichannel_2018} - is added to the mixture, with SNR (target signal to noise ratio) chosen in the interval [0 ; 20] dB. In total, network is trained with a 5-hour database, containing 1801 different rooms.

%
	
\begin{figure}
\centering
\begin{subfigure}[b]{0.14\textwidth}
\centering
\includegraphics[scale=0.25]{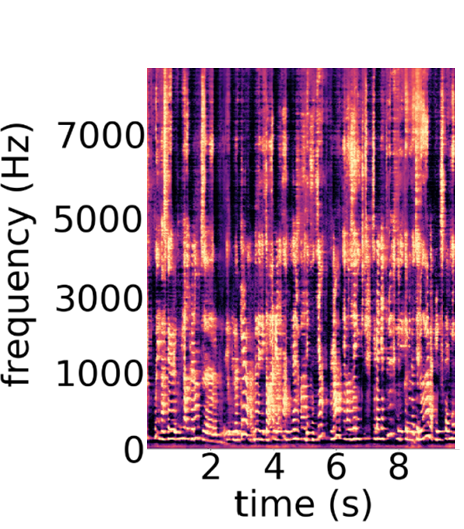}
\caption{Dilated U-net}
\end{subfigure}	
\hspace{0.1cm}
\begin{subfigure}[b]{0.14\textwidth}
\centering
\includegraphics[scale=0.25]{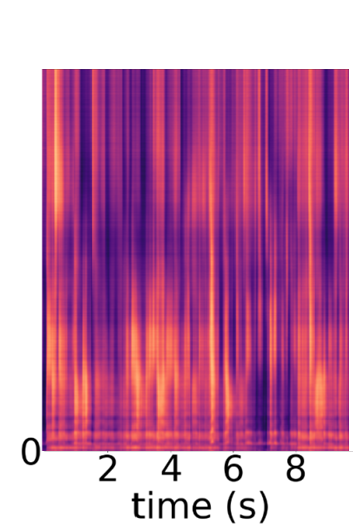}
\caption{LSTM}
\end{subfigure}	
\begin{subfigure}[b]{0.14\textwidth}
\centering
\includegraphics[scale=0.249]{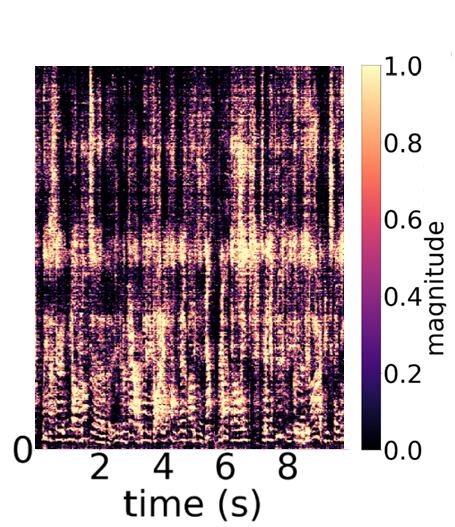}
\caption{Ideal}
\end{subfigure}	

\caption{Examples of reconstructed ratio masks.}
\label{masks}
\end{figure}

Two sets of real SRIRs are used for validation and test, with same SIR as for training and SNR set at 20~dB. Test SRIRs are recorded in a room with strong reflexions and a $\text{RT}_{60}$ around 500 ms: 32 microphone positions and 16 sources positions and orientations lead to 512 possible SRIRs.   

	The signal sampling rate is 16 kHz. We compute the Short Time Fourier Transform on 50 \% overlapped 1024-points frames, weighted by a sinusoidal window. Inputs are presented to the networks in sequences of 40 frames. The networks are trained independently for each experiment (2-3 speakers), with the least squares cost function, the Nadam optimizer, a $10^{-3}$ initial learning rate and 5\% dropout after each block. The maximum number of epochs is 50, with early stopping based on the cost function on the validation set.

	\section{Results}	
	Performance is evaluated by the word error rate (WER) metric computed on the Cobalt Speech Recognition system developed by Orange. All WERs are given with a variability of $\pm 0.5$\%, on the basis of a test we have done, computing WER 50 times on the same speech corpus at the output of the filter. WER scores are computed at the mask and filter outputs.
	U-nets (resp. LSTM) are composed of ~2 million (resp. 4 million) trainable parameters; the best model from 10 trainings is kept.

	Fig.\ref{masks} shows the accuracy of the U-nets for the mask estimation: the `fuzzy' aspect of the LSTM masks has been replaced by sharper segmentations, revealing the mask sparsity and the harmonic structure of the target spectrogram. Accordingly, the WER scores at the output of the mask (`Mask' lines in Table \ref{tab:WER}) are greatly improved: for the easiest tasks, \textit{i.e.} two speakers and 45$^{\circ}$-90$^{\circ}$, the U-nets WERs are about 50\% of the LSTM version. For the 25$^{\circ}$ case, the improvement is still noticeable with ~63\% for the dilated U-net against 77\% for the LSTM. 
	
	Scores at the output of the MWF filters show that the U-nets give the best results, regardless the configuration. For the two-speaker/90$^{\circ}$ condition - the easiest-, both U-nets reach the ideal filter score, with little but statistically significant improvement over LSTM. The gap increases with the task difficulty. For the 25$^{\circ}$ case, the WER improvement reaches 29\% (resp. 19\%) with the dilated U-net (resp. U-net). In the three-speaker case, the improvement is of same order of magnitude at about 22\% with the dilated U-net. The added value of the dilated convolution is pronounced in the difficult cases, \emph{e.g.} when the beamformer is less efficient (close sources), and when the mask sparsity is lower (two interfering speakers). In these conditions, the dilated U-net outperforms the classical U-net, almost reaching the ideal filter performance. 

	\section{Conclusion}
	In this work, we investigated U-net architectures, as a replacement for LSTM, to estimate short-term ratio masks in a speech enhancement system applied to Ambisonics contents. Tests on signals convolved by real SRIRs exhibit more precise masks, enhancing the speech harmonic structure and being more selective to interfering sources. We also show that dilated convolutions are helpful in difficult cases, including nearby sources or multiple interfering speakers, where the input features of the network appear more noisy. This is confirmed by the WER scores, where U-nets outperform LSTM in all conditions, and even reach, in the most simple cases, the performance of the oracle multichannel filter.

\newpage
\bibliographystyle{IEEEtran}
\bibliography{EUSIPCO2020_v2}


\end{document}